\documentclass[twocolumn,aps,superscriptaddress,nofootinbib]{revtex4-1}
\usepackage{latexsym,amssymb,amsmath,epsfig,bm,psfrag}
\usepackage{color}
\usepackage[bookmarksnumbered,bookmarksopen,colorlinks,citecolor=red,linkcolor=blue]{hyperref}
\usepackage{graphicx}
\usepackage{times}
\usepackage{ulem,slashed}
\usepackage{soul}
\usepackage{subfiles} 

\hypersetup{
    colorlinks=true,
    linkcolor=blue,
    filecolor=magenta,
    urlcolor=cyan,
    pdftitle={Sharelatex Example},
    bookmarks=true,
}

\newcommand{\Fig}[1]{Fig.~\ref{#1}}
\newcommand{\Eq}[1]{Eq.~(\ref{#1})}

\newcommand\be{\begin{equation}}
\newcommand\ee{\end{equation}}
\newcommand\bra{\langle}
\newcommand\ket{\rangle}

\begin{document}

\title{Extracting the Speed of Sound in Heavy-Ion Collisions: A Study of Quantum-Initiated Fluctuations and Thermalization}

\author{Yu-Shan Mu}
\affiliation{Physics Department and Center for Particle Physics and Field Theory, Fudan University, Shanghai 200438, China}

\author{Jing-An Sun}
\affiliation{Institute of Modern Physics, Fudan University, Shanghai 200433, China}
\affiliation{Department of Physics, McGill University
3600 rue University Montreal, QC Canada H3A 2T8}

\author{Li Yan}
\email{cliyan@fudan.edu.cn}
\affiliation{Institute of Modern Physics, Fudan University, Shanghai 200433, China}
\affiliation{Key Laboratory of Nuclear Physics and Ion-beam Application (MOE), Fudan University, Shanghai 200433, China}
\affiliation{Shanghai Research Center for Theoretical Nuclear Physics, National Natural Science Foundation of China and Fudan University, Shanghai 200438, China}

\author{Xu-Guang Huang}
\affiliation{Physics Department and Center for Particle Physics and Field Theory, Fudan University, Shanghai 200438, China}
\affiliation{Key Laboratory of Nuclear Physics and Ion-beam Application (MOE), Fudan University, Shanghai 200433, China}
\affiliation{Shanghai Research Center for Theoretical Nuclear Physics, National Natural Science Foundation of China and Fudan University, Shanghai 200438, China}

\begin{abstract}

The thermalization of quark-gluon plasma created in heavy-ion collisions is crucial for understanding its behavior as a relativistic fluid and the thermodynamic properties of the Quantum Chromodynamics (QCD). This study investigates the role of fluctuations in the relationship between transverse momentum and particle multiplicity, with a particular focus on their impact on extracting the QCD speed of sound. 
In a thermalized quark-gluon plasma, 
these fluctuations mostly originate from quantum fluctuations
in the colliding nuclei, and exhibit a Gaussian distribution as a consequence of their independence from thermodynamic response.
In contrast, non-thermalized systems display non-Gaussian fluctuations, reflecting the breakdown of thermalization. By leveraging the Gaussianity condition of quantum-initiated fluctuations, the physical value of the speed of sound can be extracted statistically, even in the presence of significant event-by-event fluctuations. This framework provides a robust diagnostic tool for probing thermalization and extracting thermodynamic properties in both large and small collision systems.

\end{abstract}
\maketitle

Heavy-ion collisions at facilities such as the Relativistic Heavy Ion Collider (RHIC) and the Large Hadron Collider (LHC) create extreme conditions, where Quark-Gluon Plasma (QGP) forms~\cite{Shuryak:2014zxa}. A crucial question is whether this deconfined state of quarks and gluons achieves local thermal equilibrium, at least transiently, so that one is allowed to apply hydrodynamic models~\cite{Gale:2013da,Romatschke:2017ejr,Florkowski:2017olj} and explore the phase structure of the strongly interacting matter through fundamental thermodynamic properties~\cite{Luo:2017faz}. As a closed quantum system of high energies, the thermalization of QGP is also essential for understanding non-equilibrium dynamics, in a broad sense~\cite{Schlichting:2019abc,Shen:2020mgh,Berges:2020fwq,Brauner:2022rvf}.

Local thermalization of QGP in nucleus-nucleus collisions is widely assumed, based on different types of flow signatures that reveal anisotropic particle emissions.
These are {\it indirect} evidence, however, as they are consequences of hydrodynamic response of a fluid-like QGP with respect to the geometric deformation of initial state~\cite{Heinz:2013th,Yan:2017ivm}. More explicitly, by assuming the QGP as a relativistic fluid 
in which local thermalization is implied,
the thermodynamic quantities 
like pressure $P$, energy density $e$, and entropy $S$ can be well implemented in hydrodynamic models, and the initial geometric deformation results in anisotropies of pressure gradients, which further drive the system to expand anisotropically~\cite{Ollitrault:1992bk}. 
The assumption of QGP local  thermalization has been successfully applied to
flow analyses
in central and non-central nucleus-nucleus collisions, and has been
extended to the smaller colliding systems, including proton-lead~\cite{ALICE:2012eyl,ATLAS:2013jmi,CMS:2015yux}, proton-proton~\cite{CMS:2010ifv,ATLAS:2015hzw,ALICE:2023ulm} and even electron-positron~\cite{Chen:2023njr}. 
Nonetheless, investigation of local thermalization of 
QGP  {\it directly} through the thermodynamic properties, such as the QCD speed of sound~\cite{VANHOVE1982138}, was realized only recently~\cite{Gardim:2019xjs,Gardim:2019brr}.

Among the thermodynamic properties of QCD, the speed of sound, \(c_s\), holds special significance as it contains information of the equation of state (EOS). 
Defined as 
\begin{equation}
\label{eq:cs2}
c_s^2 \equiv \frac{\partial P}{ \partial e}=\frac{\partial \ln T}{\partial \ln S}\,,
\end{equation}
the speed of sound measures the thermodynamic response of pressure to energy density, as well as the response of relative variation of temperature to the relative variation of entropy. Note that the second equation applies to a uniform and baryon-less QGP.  Speed of sound is closely linked to the stiffness of the EOS.
Near the critical region, $c_s^2$ is expected to exhibit a sharp decrease, signaling a softening of the EOS due to the transition between hadronic matter and QGP~\cite{HotQCD:2014kol}.

In high-energy ultracentral collisions of heavy nuclei, where 
geometric deformation is substantially suppressed, 
if the created QGP reaches local thermal equilibrium, 
the QCD speed of sound governs the response of variation in the observed mean transverse momentum of particles, $\bra p_T\ket\equiv \sum p_T/N_{\rm c}$, to the variation in the charged particle multiplicity, $N_{\rm c}$. This is because, in ultracentral collisions, mean transverse momentum measures the temperature in the system effectively~\cite{Gardim:2019xjs}, while the charged particle multiplicity is linked to the entropy. Consequently, linear relationships exist between $\bra p_T\ket$ and $T$, and between $N_{\rm c}$ and $S$. Corresponding to \Eq{eq:cs2},    
the response is purely thermodynamic and deterministic, so 
one has~\cite{Gardim:2019xjs,Gardim:2019brr}
\be
\label{eq:linear1}
\frac{\Delta_p}{\bra p_T\ket_0} = c_s^2 \frac{ \Delta_N}{N_0}\,,
\ee
where $\bra p_T\ket_0$ and $N_0$ are the averaged values of the mean transverse momentum and the charged particle multiplicity over the ultracentral collision events, while $\Delta_p = \bra p_T\ket - \bra p_T \ket_0$ and $\Delta_N = N_{\rm c} - N_0$ are variations. Note that the mechanism of relating $\Delta_p$ and $\Delta_N$ is not unique. From purely multi-parton interactions or mini-jet production, for instance, a linear response relation between $\Delta_p$ and $\Delta_N$ presents analogously, but the coefficient has of course, nothing to do with the speed of sound.

\begin{figure*}[t]
\begin{center}
\includegraphics[width=.90\textwidth] {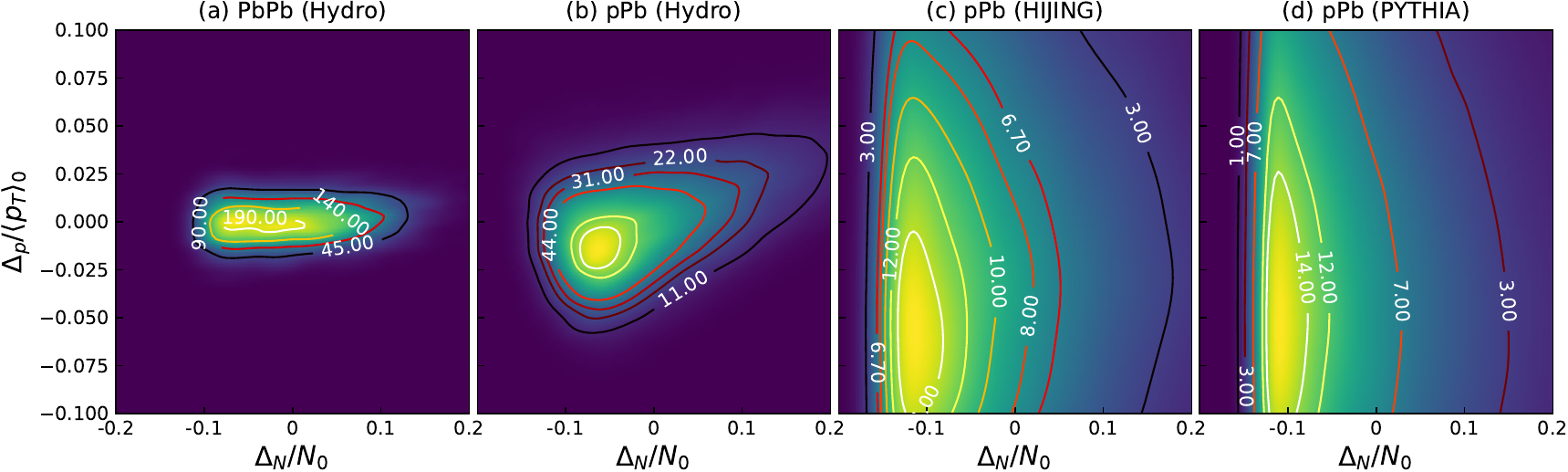}
\caption{
\label{fig:2d_dis} The normalized two-dimensional joint probability distribution of $\mathcal{P}(\Delta_N,\Delta_p)$, for the central Pb-Pb (0-5\%) and p-Pb (0-2\%) collisions at $\sqrt{s_{NN}}= 5.02$ TeV, from simulations using hydrodynamics, HIJING and PYTHIA. The tilted tip of the distributions (except PYTHIA) implies a positive correlation between $\Delta_p$ and $\Delta_N$.
}
\end{center}
\end{figure*}

While in experiments,  a linear rise between $\Delta_p$ and $\Delta_N$ has been identified in the ultracentral lead-lead collisions at $\sqrt{s_{NN}}$ = 5.02 TeV~\cite{CMS:2024sgx,ALICE-PUBLIC-2024-002}, the extracted speed of sound from the slope is contaminated. In several model studies using hydrodynamic simulations, it has been noticed that the slope depends on the selection of rapidity, cut in transverse momentum, and how centrality is determined~\cite{JFSoaresRocha:2024drz,Nijs:2023bzv}. 
These kinematic effects can be understood, at least qualitatively. For instance, 
recording in a finite interval of the transverse momentum of particle spectrum reduces the number of thermal particles taken into account, which as a consequence changes the contribution of thermodynamic response. 

On the other hand, there are fluctuations in realistic heavy-ion collisions. On an event-by-event basis, these fluctuations are dominantly of quantum origin, such as the fluctuations of nucleons populating inside nuclei as they collide, and decay and scatterings of hadrons during final-state particlization~\cite{thermalfluct}.
In terms of flow measurement, fluctuations in high-energy heavy-ion collisions have been well captured through multi-particle correlations~\cite{Shen:2020mgh}, consistent with hydrodynamic model simulations on an event-by-event basis.
The role of fluctuations in the response relation between $\Delta_p$ and $\Delta_N$,
in the ultracentral collisions, remains unclear so far~\cite{Sun:2024zsy}.  In this Letter, we will investigate the effect of fluctuations on the extraction of speed of sound,  and we propose a statistical method that systematically isolates the thermodynamic response, so that in both theoretical simulations and experiments the speed of sound can be well identified in the presence of the fluctuations. More importantly, 
we find that 
these quantum-initiated 
event-by-event fluctuations~\cite{quantum} make the extraction of the speed of sound an ideal probe of the QGP thermalization.

{\it Quantum-initiated fluctuations.} --
In realistic heavy-ion collisions, 
with respect to the mean transverse momentum and the charged particle multiplicity,
the presence of fluctuations leads to a two-dimensional joint probability distribution of $\mathcal{P}(\Delta_p,\Delta_N)$~\cite{Giacalone:2020lbm,Samanta:2023amp,Samanta:2023kfk,ATLAS:2024jvf}, instead of the deterministic relation \Eq{eq:linear1}. 

While such a distribution can be measured experimentally, for the purpose of illustration, we generate a set of $\mathcal{P}(\Delta_p,\Delta_N)$ from numerical simulations of hydrodynamic model and non-thermal models~\cite{numberofevents}. The hydrodynamic model is hybrid that incorporates the generation of initial density profile using the T$_{\rm R}$ENTo method~\cite{Trento2Moreland:2014oya}, solving coupled equations of viscous hydrodynamics with the EOS from lattice QCD (LEOS)~\cite{HotQCD:2014kol} for the system's dynamical evolution using the MUSIC program~\cite{hydroSchenke:2010rr,hydroSchenke:2010nt,hydroPaquet:2015lta,hydroparam}, which is then connected to UrQMD for particlization and hadron gas evolution~\cite{urqmdBass:1998ca}.  
By construction, the quantum-initiated fluctuations that are dominantly from the initial state are accounted for stochastically in the hybrid modeling.
It should be emphasized that while solving hydrodynamics implies local thermalization up to the dissipative corrections, the hadron dynamics regarding UrQMD does not require local thermalization. Therefore, local thermalization 
in the hybrid hydrodynamic model is only transiently realized.

For lead-lead (Pb-Pb) collisions at high energies, longitudinal boost invariance is a good approximation, which allows us to simplify the analysis and work in a 2+1 dimensional setup. However, this longitudinal symmetry is broken in asymmetric systems such as proton-lead (p-Pb), for which we invoke the exact 3+1 dimensional characterization of the initial state in T$_{\rm R}$ENTo-3D~\cite{Soeder:2023vdn}, as well as in MUSIC. For comparison, we also simulate and collect data on the mean transverse momentum and multiplicity in p-Pb collisions using non-thermal models, including HIJING~\cite{HijingWang:1991hta,HijingDeng:2010mv,Wang:1991us} and PYTHIA~\cite{PythiaSjostrand:2007gs,PythiaSjostrand:2014zea,PythiaBierlich:2022pfr}. In these non-thermal models, particles are generated by parton scatterings and radiative processes rather than thermal medium expansion, 
without 
local thermalization. 

\begin{figure*}[t]
\begin{center}
\includegraphics[width=1.0\textwidth] {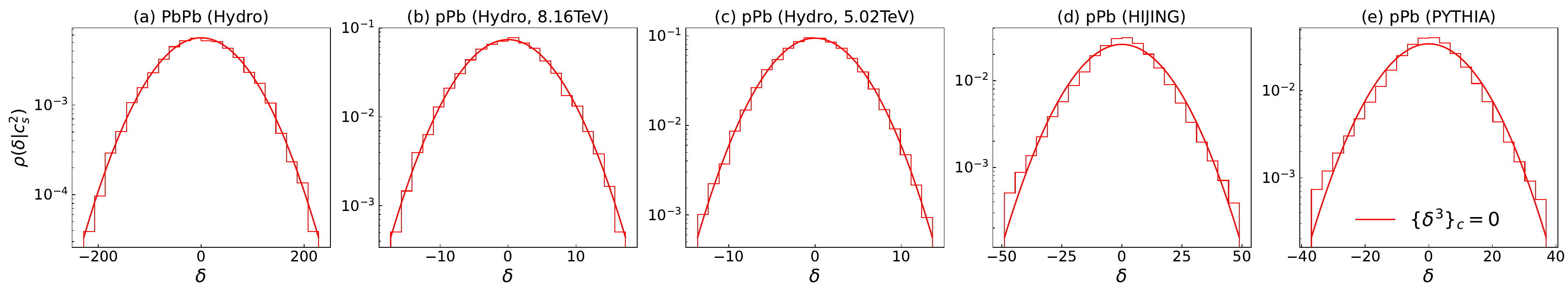}
\caption{
\label{fig:gauss} The normalized probability distribution $\rho(\delta|c_s^2)$,  with $c_s^2$ determined by the zero skewness condition $\{\delta^3\}_c=0$ (red solid histogram). 
For comparison, the Gaussian distributions with zero mean and the variance $\{\delta^2\}^{1/2}$ determined with respect to the extracted $c_s^2$ values are shown by red solid line.
}
\end{center}
\end{figure*}

With 
the event averaged values of $\bra p_T\ket$ and $N_c$ fixed with respect to the 
top 5\% centralities for Pb-Pb collisions 
and top 2\% centralities for p-Pb collisions
at the LHC energies, 
for particles of pseudo-rapidity $|\eta|<0.5$ and transverse momentum $p_T\in[0,3]$ GeV~\cite{ATLAS:2015hkr,ALICE:2022xip,CMS:2015yux}, from numerical simulations we obtain the two-dimensional joint probability distributions, as presented in \Fig{fig:2d_dis}.
Note that from the hydrodynamic simulations of central Pb-Pb collisions (panel (a)), to HIJING and PYTHIA simulations of central p-Pb collisions (panel (c) and (d)), the width of these distributions grows significantly, indicating increasing fluctuation strength.
In Pb-Pb collisions, the vanishing impact parameter, or saturation of system size, appears only in the ultra-central events, namely, the right tip of the distribution  in \Fig{fig:2d_dis} (a)~\cite{Samanta:2023kfk,Gardim:2019brr}.
In p-Pb collisions, although initial geometry is dominantly determined by the proton, the effective system size saturates only in the ultra-central collisions~\cite{Gardim:2022yds}. 
Therefore, in the following for the extraction of speed of sound, we only focus on the $0$-$2.5\%$ centrality class of Pb-Pb collisions and 0-1\% centrality class in p-Pb collisions.
For later convenience, we also obtained the effective temperatures from the freeze-out hypersurface~\cite{Gardim:2019xjs,Supple},. 

Given the joint probability distribution $\mathcal{P}(\Delta_p, \Delta_N)$, the response between
the variation of the transverse momentum and the variation of multiplicity should be generalized with 
a stochastic contribution, $\delta$,
\begin{equation}
\label{eq:master}
\frac{\Delta_p}{\bra p_T\ket_0} = c_s^2 \frac{ \Delta_N + \delta}{N_0} \,.
\end{equation}
Importantly, for a thermalized QGP, $\delta$ is independent of the thermodynamic response relation. This independence reflects the quantum origin of $\delta$, which introduces stochastic fluctuations uncorrelated with the deterministic relationship. While 
$\delta$ is independent of the thermodynamic response, it may exhibit statistical correlations with individual observables like 
$\Delta_N$ or $\Delta_p$, arising from shared quantum or initial-state fluctuations. In a thermalized system, where $\delta$ varies randomly and independently across events, its distribution approaches Gaussianity by the Central Limit Theorem (CLT)~\cite{Samanta:2023amp,Samanta:2023kfk}. 
On the other hand, in a system without thermalization, the response relation itself (where the response coefficient 
is no longer the speed of sound) is quantum, hence $\delta$ is not independent from event to event and the $\delta$-distribution is generically non-Gaussian.
This non-Gaussianity in $\delta$ in non-thermal systems can be conceptually understood as a consequence of interactions among these fluctuations, through the quantum response relation. This is analogous to the non-Gaussianity observed in primordial quantum fluctuations, which signals interactions in the early universe~\cite{Maldacena:2002vr}.

Irrespective of the condition of local thermalization, finite-size effects, proportional to $1/N_0$, introduce residual non-Gaussianity to the $\delta$-distribution. Specifically, corrections to higher-order moments of the $\delta$-distribution scale as powers of $1/N_0$, with the skewness scaling as 
$1/N_0^{3/2}$, and kurtosis scaling as $1/N_0^2$, etc. Thus, the previous discussions of Gaussianity in thermalized QGP and non-Gaussianity in non-thermal systems apply up to corrections of order $1/N_0^a$, where $a$ depends on the specific observable or moment. These finite-size effects are particularly relevant 
when $N_0$ is relatively small. 

To verify the analysis, we need to solve the distribution of $\delta$ with respect to the physical value of the speed of sound. Regarding \Eq{eq:master}, 
it is a mathematical problem:  How to solve the constant parameter $c_s^2$, given two known distributions of $\Delta_p$ and $\Delta_N$, and one unknown distribution of $\delta$? 
 In fact, 
 \Eq{eq:master} implicitly defines the conditional distribution, $\rho(\delta | c_s^2)$. In the small $c_s^2$ limit, $\rho(\delta | c_s^2)$ approaches the distribution of $\Delta_p/\bra p_T\ket_0$, while for $c_s^2\to \infty$, the distribution of $\delta$ is equivalent to that of $-\Delta_N/N_0$.

One first notices a trivial relation, by definition, in the ultracentral collisions, $\{\delta\}= \frac{1}{N_{\rm events}}\sum_{\rm events} \delta=0$,
where the brackets $\{\ldots\}$ denote average over the ultracentral events.
For a  thermalized QGP, given the joint probability distribution $\mathcal{P}(\Delta_p,\Delta_N)$ with the physical value of the speed of sound, the 
fluctuation $\delta$ is Gaussian. Consequently, the physical value of the speed of sound implies vanishing cumulants of the $\delta$-distribution of arbitrary orders, up to corrections of $1/N_0^a$. To proceed, the simplest calculation is to solve the zero skewness condition, $\{\delta^3\}_c=\{\delta^3\}=0$, and then we examine the Gaussianity of the $\delta$-distribution accordingly. 
The zero skewness condition yields an equation of $c_s^2$, namely,
\be
\label{eq:d3}
(c_s^2)^3 \frac{\{\Delta_N^3\}}{N_0^3} 
- 3 (c_s^2)^2 \frac{\{\Delta_N^2 \Delta_p\}}{N_0^2 \bra p_T\ket_0}
+ 3 c_s^2 \frac{\{\Delta_N \Delta_p^2\}}{N_0  \bra p_T\ket_0^2} - \frac{\{\Delta_p^3\}}{\bra p_T\ket_0^3} = 0\,,
\ee
with the physical value of the speed of sound its real root. The coefficients of the equation correspond respectively, to the standardized skewness or mixed skewness of the joint probability distribution $\mathcal{P}(\Delta_p, \Delta_N)$, which are all measurable in realistic experiments. 
With the speed of sound extracted from the zero skewness condition, one is allowed to solve the exact probability distribution $\rho(\delta|c_s^2)$. As shown in \Fig{fig:gauss}, to a good approximation within corrections of $1/N_0^a$, we find that $\delta$ from hydrodynamic simulations follow the Gaussian distribution, while from non-thermal simulations, fluctuations of $\delta$ exhibit apparently non-Gaussianity with heavy tails. We have also solved similarly, with respect to the condition that the fifth order cumulant $\{\delta^5\}_c=\{\delta^5\}-10\{\delta^2\}\{\delta^3\}=0$, 
which gives consistent results~\cite{Supple}.
Besides, for the values of $c_s^2$ other than the desired ones, we notice that even from hydrodynamic models $\rho(\delta|c_s^2)$ becomes non-Gaussian.

{\it Extracting speed of sound in the presence of fluctuations.} --
Without any concern of the effect of fluctuations, \Eq{eq:master} can be solved via a sub-bin selection of the ultracentral events, as has been carried out so far in experiments~\cite{CMS:2024sgx,ALICE-PUBLIC-2024-002} and theoretical model studies~\cite{JFSoaresRocha:2024drz,Nijs:2023bzv}. After event average in the sub-bins, the variation of the mean transverse momentum and multiplicity can be quantified between sub-bins, and one has 
\be
\frac{\{\Delta_p\}_I}{\bra p_T\ket_0}  = c_s^2 \frac{ \{\Delta_N\}_I+ \{\delta\}_I}{N_0} \,,\quad I = 1,2,\ldots
\ee 
where $\{\ldots\}_I$ denotes event average in the sub-bin $I$. If $\delta$ and $\Delta_N$ are independent in the sub-bins, the speed of sound can be simply read off as the slope from the resulted line between $\{\Delta_p\}_I/\bra p_T\ket_0$ and $\{\Delta_N\}_I/N_0$, otherwise, the slope gets correction from $\{ \delta\}_I\ne 0$. 
In fact, the condition $\{\delta\}=0$ implies already in the sub-bins, $\{\delta\}_I = \alpha \{\Delta_N\}_I/N_0$, with $\alpha$ a constant. One may understand this linear relation as the leading order expansion of $\{\delta\}_I$ in terms of $\{\Delta_N\}_I/N_0$.

\begin{figure}[t]
\begin{center}
\includegraphics[width=.45\textwidth] {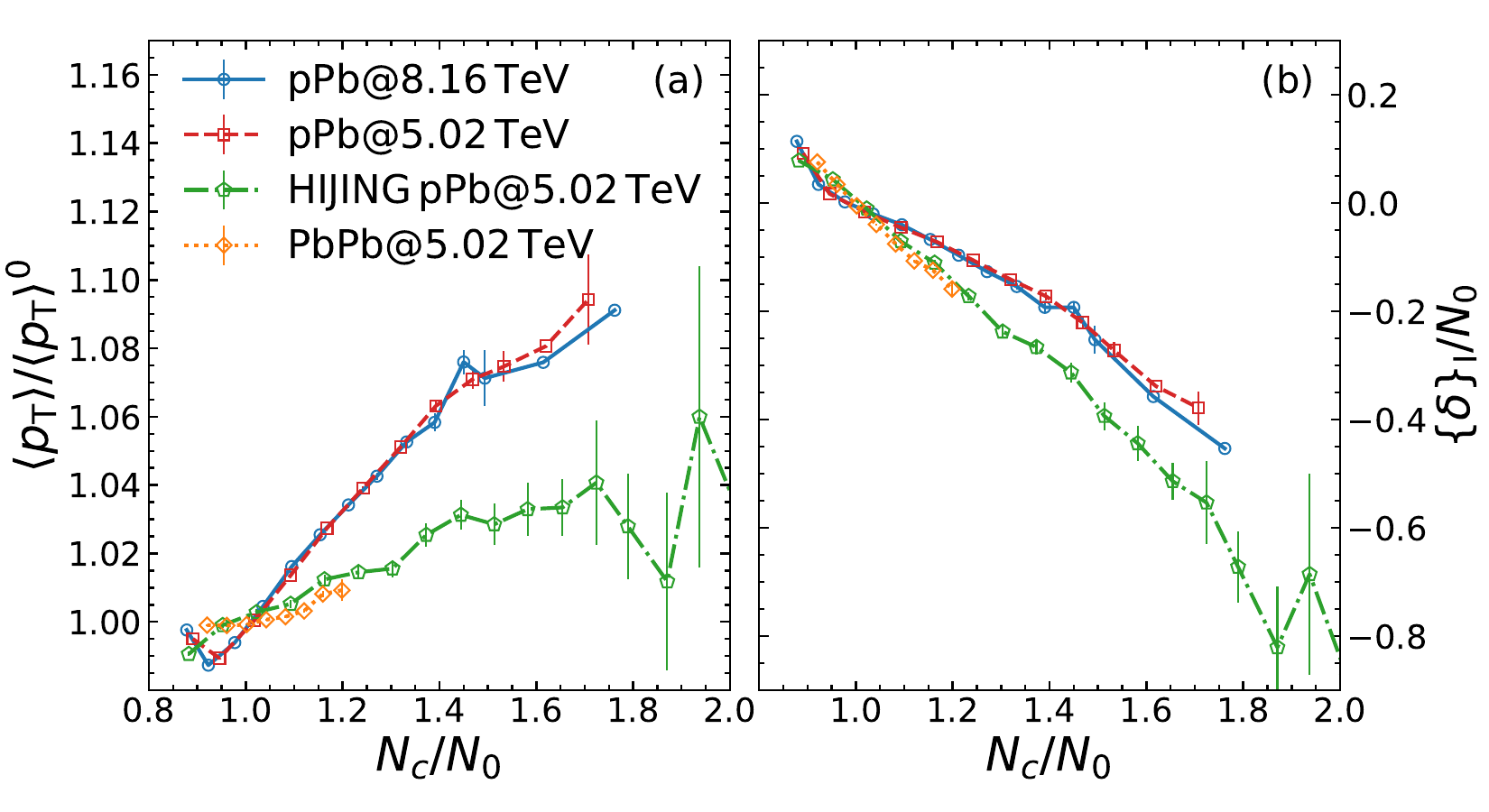} 
\caption{
\label{fig:linear} (a) Linear correlation between $\{\bra p_T\ket_I\}$ and $\{N_c\}_I$ from sub-bins of the central events from hydrodynamic and HIJING models. (b) Negative linear correlation between $\{\delta\}$ and $\Delta_N$ shown in the sub-bin averages of $\{\delta\}_I$. 
}
\end{center}
\end{figure}

\begin{table*}[] 

\begin{tabular}{c|l|l|l|l|l}
\hline
$c_s^2$   & sub-bin slope & $\{\delta^3\}_c=0$ & $\{\delta^5\}_c=0$ & LEOS  & $T_{\rm eff}$ (MeV)\\ \hline
PbPb (Hydro, 5.02TeV)&$0.123 \pm 0.035 $& $0.217\pm0.032$ & $0.216\pm0.041$  &   0.222-0.242 &  210.3  \\ 
\hline
pPb (Hydro, 8.16 TeV) & $0.176\pm0.004$ & $0.292\pm0.013$ & $0.287\pm0.012$ &   0.282-0.309   & 319.9-382.1\\ 
pPb (Hydro, 5.02 TeV)& $0.197\pm0.004$ & $0.318\pm0.011$ & $0.313\pm0.008$&  0.269-0.304  &280.2-353.4\\ 
pPb (PYTHIA, 5.02 TeV)& $-0.032\pm0.002$&$1.178\pm0.006$  & $1.352\pm0.019$ &    0.227-0.278 & 203.2-256.4 \\ 
pPb (HIJING, 5.02 TeV) &$0.079\pm0.003$&$1.104\pm0.019$  & $1.171\pm0.053$ &    0.206-0.271  & 219.8-277.2\\ 
\hline
\end{tabular}
\caption{\label{table:sum}
A summary of the extracted speed of sound from different methods for various model simulations. 
By determining the effective temperatures via their linear relation with $\bra p_T\ket_0$, $c_s^2$ from LEOS can be estimated~\cite{HotQCD:2014kol}. Note that unlike in PbPb collisions where one approximately has $\bra p_T\ket \approx 3T_{\rm eff}$, in pPb collisions longitudinal expansion introduces larger uncertainty in the proportionality coefficient.
}
\end{table*}

In \Fig{fig:linear} (a), using the sub-bin method, the linear correlations between $\{\bra p_T \ket\}_I/\bra p_T\ket_0$ and $\{N_{\rm c}\}_I/N_0$ 
from the different models and systems are shown. Except the results from PYTHIA, where the linear relation has a negative slope, all the model simulations indicate a positive linear correlation between $\Delta_p$ and $\Delta_N$. Notably, from the HIJING model, although there is no local thermalization, one still finds a slope that is sizable in magnitude. However, even from hydrodynamic simulations, we find that the values of the slope are substantially small comparing to the physical values of $c_s^2$ from the LEOS expectations, suggesting a finite and negative contribution from $\{\delta \}_I$. 
Indeed, as demonstrated in \Fig{fig:linear} (b), where the averaged fluctuations in each sub-bin is plotted versus $N_c/N_0$, a clear negative correlation emerges between $\delta$ and $N_c$, reflecting the suppression of quantum fluctuations by larger system entropy.

Corresponding to the slope of the sub-bin linearity, and the 
statistical method using
Gaussianity condition, 
we solve the physical values of the speed of sound. A summary of the obtained results is given in Table.~\ref{table:sum}. We notice that the extracted values of the speed of sound for all the hydrodynamic model simulations from the Gaussian distribution condition are compatible to the LEOS expectations, within uncertainties. For the non-thermal model simulations, however, the same procedure yields non-physical values, as the speed of sound appears greater than unity (speed of light  $c=1$ as we take the natural units), violating the causality condition~\cite{Cherman:2009tw,Hippert:2024hum}.  
Furthermore, since the extraction of \( c_s^2 \) via the linear slopes of \( \Delta_p \) and \( \Delta_N \) is systematically biased by the negative correlation between 
$\delta$ and \( N_c \), 
determining \( c_s^2 \) within the Gaussianity framework is more reliable where such correlations are inherently accounted for.


{\it Probe of thermalization.}--
While the exact value of the speed of sound is of great significance to QCD dynamics, we emphasize that the process of extracting the speed of sound also serves as a direct probe of the thermalization of the QGP, which is also important. The response between mean transverse momentum and particle yield fundamentally differs in a locally thermalized system compared to a non-thermal system. These differences can be effectively distinguished by analyzing the 
statistical properties
 of $\delta$ on an event-by-event basis.

Based on our findings, we propose a combined measurement of the speed of sound and the Gaussianity of $\delta$ to characterize local thermalization. In a thermalized system, one expects the extracted speed of sound to have a physical value that falls within a reasonable range~\cite{Cherman:2009tw,Hippert:2024hum}, despite uncertainties such as those arising from the determination of the effective temperature, $T_{\rm eff}$, which is partially model-dependent. Simultaneously, the probability distribution $\rho(\delta| c_s^2)$ should exhibit Gaussianity, up to corrections of order $1/N_0^a$. In contrast, if thermalization is not achieved, neither a physical value of the speed of sound nor Gaussianity in $\delta$ should be observed.

Moreover, deviations from thermalization can be quantified using higher-order cumulants of the $\rho(\delta| c_s^2)$ distribution. For example, the standardized kurtosis, defined as $\kappa_4 = \{\delta^4\}/\{\delta^2\}^2-3$, should vanish in a system with complete local thermalization but remain finite in the presence of non-thermal contributions. From our numerical simulations, we find 
 $\kappa_4 =2.15\pm 0.05$ and $1.91 \pm 0.11$ in HIJING and PYTHIA models, respectively. In contrast, hybrid hydrodynamic modeling yields 
 $\kappa_4=0.16\pm 0.05$ and $0.18\pm 0.07$ for p-Pb collisions at  $\sqrt{s_{NN}}=5.02$ TeV and $8.16$ TeV, and $\kappa_4=-0.064 \pm 0.096$ for Pb-Pb collisions, signaling small but finite non-thermal contributions.

{\it Summary and discussion.} --
For a QGP system created in ultracentral nucleus-nucleus collisions, if it is locally thermalized, the response relation between $\Delta_p$ and $\Delta_N$ is thermodynamic. Accordingly, event-by-event fluctuations, which are primarily quantum in origin, should be independent of the thermodynamic response and follow a Gaussian distribution. Although these fluctuations are correlated with $\Delta_N$ and induce corrections to averages, the Gaussianity condition of $\delta$ allows for the statistical extraction of the physical value of the speed of sound, for example, by solving the equation derived from the condition $\{\delta^3\}_c=0$.
We expect the coefficients in this equation to be experimentally measurable, as they correspond to the standardized skewness or mixed skewness of the joint probability distribution $\mathcal{P}(\Delta_N,\Delta_p)$. Therefore, the physical value of the speed of sound can be determined in experiments, even in systems where fluctuations are substantial, such as in small colliding systems like p-Pb. 
Note however, the inferred speed of sound is subject to uncertainties arising from initial-state modeling and viscous corrections~\cite{hydroparam}, which warrant further systematic investigation through dedicated theoretical studies.

The non-Gaussianity of $\delta$ can serve as a robust diagnostic tool for thermalization in QGP systems. 
When $\kappa_4$ of $\delta$ exceeds unity, the dominant contribution to the response between $\Delta_p$ and $\Delta_N$ is no longer thermodynamic, indicating that the system should not be considered thermalized.
Noting that a deterministic thermodynamic response involving $c_s^2$ remains valid in the presence of finite baryon density, the current formulation is expected applicable
at lower collision energies~\cite{nextpaper}, where the characterization of thermalization becomes even more crucial for the search of the QCD critical point.

{\it Acknowledgements.} --
We are grateful to Jean-Yves Ollitrault, Wei Li, and Jiangyong Jia for very helpful discussions.  We also thank Xin-Nian Wang for suggesting the HIJING model and Xiangyu Wu for the HIJING code support.
Y.-S. M. would like to thank Chen Zhong, Lu-Meng Liu and Wen-Hao Zhou for their assistance with the computation server. This work is supported in part by the National Natural Science Foundation of China through Grants No. 12375133 and No. 12147101 (L.Y.), No. 12147101, No. 12225502, and No. 12075061 (X.-G. H.), by the Natural Science Foundation of Shanghai through Grant No. 23JC1400200, and by the National Key Research and Development Program of China through Grant No. 2022YFA1604900. The computations are performed at the CFFF platform of Fudan University.

\bibliography{references}

\begin{thebibliography}{56}
\bibitem [{the()}]{thermalfluct}%
  \BibitemOpen
  \href@noop {} {}\bibinfo {note} {There are thermal fluctuations while the QGP
  stays close to local thermal equilibrium, whose effect is essentially
  suppressed as it is proportional to the inverse of the specific heat,
  $c_v^{-1}=c_s^2/S$.}\BibitemShut {Stop}%
\bibitem [{num()}]{numberofevents}%
  \BibitemOpen
  \href@noop {} {}\bibinfo {note} {We accumulate approximately 5,000 central
  events for Pb-Pb collisions and 20,000 central events for p-Pb collisions
  from hydrodynamic simulations. In addition, we obtain around one million
  central collision events from simulations using HIJING and
  PYTHIA.}\BibitemShut {Stop}%
\bibitem [{hyd()}]{hydroparam}%
  \BibitemOpen
  \href@noop {} {}\bibinfo {note} {In our current hybrid hydrodynamic model
  simulations, we used a standard set of parameters, including $\eta/s=0.2$,
  $\zeta/s=0$, and fluid kinetic freeze-out energy density at $e_{\rm fo}=0.18$
  GeV/fm$^3$.}\BibitemShut {Stop}%
\end{thebibliography}


\end{document}